\documentclass[]{article}
\usepackage{amsmath,amssymb,comment}
\usepackage{cite}
\usepackage{graphicx}
\usepackage[margin=1in]{geometry} % remove, eventually

\usepackage{mathtools}

\newtheorem{theorem}{Theorem}
\newtheorem{corollary}{Corollary}
\newtheorem{lemma}{Lemma}

\DeclareMathOperator{\diag}{diag}

\DeclareMathOperator{\supp}{supp}
\DeclareMathOperator{\Tr}{Tr}
\DeclareMathOperator*{\argmin}{\arg\min}

\newcommand{\overbar}[1]{\mkern 3mu\overline{\mkern-3mu#1\mkern-1.5mu}\mkern 1.5mu}

\makeatletter
\if@twocolumn
\newcommand{\iftwocol}[1]{#1} % only add in twocolumn
\newcommand{\ifnottwocol}[1]{} % don't add in twocolumn
\newcommand{\tfractwocol}[2]{\tfrac{#1}{#2}} % use tfrac if two column
\else
\newcommand{\iftwocol}[1]{} % else nothing
\newcommand{\ifnottwocol}[1]{#1} % don't add in twocolumn
\newcommand{\tfractwocol}[2]{\frac{#1}{#2}} % else use \frac
\fi
\makeatother

\title{Estimation of Poisson arrival processes under linear models}
\author{Michael~G.~Moore and Mark~A.~Davenport\thanks{M.~Moore and M.~Davenport are with the Georgia Institute of Technology in Atlanta, GA, USA.}
\iftwocol{\thanks{This work was supported by grants NRL N00173-14-2-C001, AFOSR FA9550-14-1-0342, and NSF CCF-1350616 as well as support from the Alfred P. Sloan Foundation.}}
}

\begin{document}
\maketitle

\begin{abstract}
In this paper we consider the problem of estimating the parameters of a Poisson arrival process where the intensity function is assumed to lie in the span of a known basis.  Our goal is to estimate the basis expansions coefficients given a realization of this process.  We establish novel guarantees concerning the accuracy achieved by the maximum likelihood estimate. Our initial result is near-optimal, with the exception of an undesirable dependence on the dynamic range of the intensity function. We then show how to remove this dependence through a process of ``noise regularization'', which results in an improved bound under our analysis. We conjecture that a similar guarantee should be possible when using a more direct (deterministic) regularization scheme. We conclude with a discussion of practical applications and an empirical examination of the proposed regularization schemes.
\end{abstract}

\section{Introduction}

In a wide range of practical problems, we wish to understand an underlying process given observations consisting only of the times (or locations) of certain events of interest. Such problems arise in as diverse applications as fluoroscopy \cite{singh_enhancement_1993} and mass spectrometry \cite{cotter_time-of-flight_1993}, or when modeling earthquakes \cite{ogata_space-time_1998}, neurons \cite{johnson_point_1996}, cell tissue \cite{lagrow_approximating_2018}, social networks \cite{zhou_learning_2013-1}, or terrorism incidents \cite{tench_spatio-temporal_2016}, just to name a few.
In this paper we consider the case where the events are generated according to a linear parametric Poisson observation model. Specifically, we assume the events are generated according to some unknown intensity function and aim to estimate the parameterization of the intensity in a known basis.  Our goal is to understand how various characteristics of the generative model relate to the quality of the estimated parameters.

To be more precise, we will assume that we observe events drawn from a Poisson arrival process over a Lebesgue-measurable space $\mathbb{T}$\footnote{We will typically think of $\mathbb{T}$ as simply representing an interval of $\mathbb{R}$, but leave this general to allow for more complex (e.g., spatio-temporal) Poisson processes, to which our analysis also applies.}
with measure $|\mathbb{T}|$ corresponding to the volume of this space.
We will consider a set of known basis functions $g: \mathbb{T}\to\mathbb{R}$ and $\gamma_n: \mathbb{T}\to\mathbb{R}$ and unknown coefficients $\bar{x}_n\in\mathbb{R}$ for $n = 1, \ldots, N$ that define a Poisson intensity function $R_{\bar{x}}:\mathbb{T}\to\mathbb{R}_+$ of the form
\begin{equation} \label{eq:rate}
R_{\bar{x}}(t) = g(t) + \sum_{n=1}^N \bar{x}_n \gamma_n(t) = g(t) + \bar{x}^T \gamma(t),
\end{equation}
where $\gamma(t) = [\gamma_1(t), \ldots, \gamma_N(t)]^T$ and $\bar{x} = [\bar{x}_1, \ldots, \bar{x}_N]^T$. Our observations consist of event coordinates $\tau_1, \ldots, \tau_M \in \mathbb{T}$ (or $\tau = \{\tau_m\}_{m=1}^M$ for short) drawn according to this process, by which we mean that the event coordinates satisfy
\begin{equation} \label{eq:arrivalmodel}
|\tau \cap T| \sim \text{Poisson}\left(\int_T R_{\bar{x}}(t)dt\right)
\end{equation}
for any $T\subseteq\mathbb{T}$, where here $|\cdot|$ denotes the cardinality of a set.  We note that the total number of events we observe is $M=|\tau|$, which is a Poisson-distributed random variable with parameter $\overbar{M}=\mathbb{E}(M)=\int_{\mathbb{T}} R_{\bar{x}}(t)dt$.  We also note that, conditioned on $M$, each $\tau_m$ is independent and identically distributed with probability density function $f_{\tau_m}(t) = \frac{1}{M} R_{\bar{x}}(t)$.
Given the observed events $\tau$, our goal is to estimate the vector $\bar{x}$.

A primary motivation for this work is to obtain a result that applies to arbitrary sets of basis functions $\gamma_n(t)$.  Existing work often focuses on bases with desirable properties.  While it is sometimes possible for the user to carefully construct a basis with desirable properties that result in a simpler analysis, in many important cases the basis is determined by the application.
For example, consider the following applications:
\begin{itemize}
\item \textbf{Low-light imaging:} Suppose that we desire to resolve a sparse image from blurry photon counts~\cite{lo_nonlinear_1979}. In this case the natural basis is composed of shifted versions of the point spread function inherent to the system, and the coefficients in this basis represent the resolved pixel values we wish to recover.
\item \textbf{Oversampled mass spectrometry:}  The mass spectrum of an ionized substance (the mass-to-charge ratios of the constituent ions) can be measured by applying an electrical impulse to the ions and measuring their corresponding velocity.  This is accomplished by sensing when they hit a detector after traveling across a distance.  If the impulse is applied rapidly to the substance~\cite{moore_randomized_2015}, the observations at the detector is the mass spectrum convolved with the electrical impulse pattern.  In this setting the basis elements are shifted (and sometimes blurred) versions of the impulse pattern, one for each mass-to-charge ratio under consideration (which determines the size of the shift).  The coefficients estimated using this basis then determine the mass spectrum of the sample.
\item \textbf{Autoregressive Poisson processes:} Linear autoregressive Poisson processes \cite{fokianos_poisson_2009,hall_inferring_2016} or Hawkes processes \cite{hawkes_point_1971,tench_spatio-temporal_2016,moore_analysis_2016} correspond to networks of mutually-exciting temporal point processes.  When attempting to estimate the network parameters of these processes, we consider a Poisson-like (though, in general, not Poisson) estimation problem where the basis is causally constructed as a function of previous events.  Specifically, the basis is constructed by convolving a known causal kernel with the preceding event times.  Under certain assumptions on the structure of a network (e.g., that the network forms a directed acyclic graph), these observations correspond to observations of an ensemble of Poisson processes with (random) bases. The coefficients in this basis then determine the dictate the degree of influence of the other processes in the network on the process under consideration.
\end{itemize}

In each of these examples, we have almost no influence over the basis selection process but are interested in the basis representation of the process.  We must be capable of analyzing the performance of our processing for a wide variety of bases.  Towards this end, the results we present in this paper are based on simple basis properties (e.g., norms on $\gamma_n(t)$ and properties related to the conditioning of the basis), which make them particularly suited to these situations where we do exercise direct control over the basis.

\subsection{Main contribution}

In Theorem~\ref{thm:main} we provide an error bound for the maximum-likelihood estimate of $\bar{x}$ in the linear model of~\eqref{eq:rate}. To the best of our knowledge, this is the first such guarantee for this setting
(with existing results applying principally to the case where $\mathbb{T}$ is a discrete set, as discussed in Section~\ref{sec:relatedwork}).
The result roughly states that if $R_{\text{min}} \le R_{\bar{x}}(t) \le R_{\text{max}}$ where $R_{\text{min}}$ is sufficiently large, then with high probability the maximum likelihood estimate $\widehat{x}$ will satisfy
\begin{equation} \notag
\|\widehat{x}-\bar{x}\|_2 \le C_{\gamma} \frac{R_{\text{max}}}{\sqrt{R_{\text{min}}}},
\end{equation}
where $C_{\gamma}$ is a constant that depends on the conditioning of the basis $\gamma(t)$.  In Theorem \ref{thm:sparserec}, we extend this result to the special case where $\bar{x}$ and the estimate are known to be sparse, although in practice this may require that we solve a combinatorial optimization.

We observe that our performance bound exhibits no dependence on the $g(t)$ term in~\eqref{eq:rate}. Thus, it is possible to artificially reduce the dynamic range $\frac{R_{\text{max}}}{R_{\text{min}}}$ to a constant by augmenting the observations $\tau$ with additional events generated by a homogenous Poisson process (thereby uniformly increasing $g(t)$ and reducing the dynamic range). This ``noise augmentation'' procedure can be viewed as a form of regularization, which allows us to establish an improved bound (Corollary~\ref{thm:noised}) of the form
\begin{equation} \notag
\|\widehat{x}-\bar{x}\|_2 \le C_{\gamma}' \sqrt{R_{\text{max}}}
\end{equation}
with no requirement on $R_\text{min}$.
Our result is, to the best of our knowledge, the first to remove this dependence on the dynamic range.  Previously, this dependence limited the situations to-which theoretical results were applicable.

We conjecture that a bound of this form should also apply to an alternative (deterministic) regularization scheme, described in Section~\ref{sec:noisebound}.  Note, however, that in Section~\ref{sec:sim} we empirically study the impact of these regularization schemes and find little evidence for improvements in practice. This suggests that the dependence on the dynamic range in Theorem~\ref{thm:main} may simply be an artifact of our proof and the proofs of related results.

Finally, we also compare our results to estimates of the Cram\'{e}r-Rao lower bound, as well as to the error we expect in a simple concrete choice of basis $\gamma(t)$. Together, these comparisons suggest that when the basis is relatively well-conditioned, there is little potential room for improvement in our bounds (up to a constant and the dynamic range term in Theorem \ref{thm:main}). On the other hand, there may still be room for improvement when the basis is poorly conditioned.

\subsection{Related work} \label{sec:relatedwork}

Our focus in this paper is on the general Poisson arrival model of~\eqref{eq:arrivalmodel}. This model has received relatively little attention up to this point. It is considered in~\cite{hansen_lasso_2015}, which studies a LASSO-based estimator, but which provides bounds only on the accuracy with which $R_{\bar{x}}(t)$ is recovered, rather than $\bar{x}$ as is our focus here. Far more common in the literature is a related counting-based model which is used in, e.g., \cite{li_minimax_2016,moore_randomized_2015,rohban_minimax_2016,jiang_data-dependent_2015,raginsky_compressed_2010,soni_estimation_2014}. In this model we instead observe Poisson bin counts in the form
\begin{equation} \label{eq:countingmodel}
y_m \sim \text{Poisson}\left(g_m + \langle\gamma_m,\bar{x}\rangle\right)
\end{equation}
for $m = 1, \ldots, M_0$ (we retain $M=\sum_m y_m$ as the number of events).  This can be viewed as a discrete approximation to the arrival model, realized by partitioning $\mathbb{T}$ into non-overlapping intervals $T_m$ and setting \label{ref:countingsub} $y_m = |\tau\cap T_m|$, $g_m = \int_{T_m} g(t) dt$, and $\gamma_m = \int_{T_m} \gamma(t)dt$.
In fact, the arrival model in \eqref{eq:arrivalmodel} is identical to the counting model in \eqref{eq:countingmodel} in the special case that $g(t)$ and $\gamma_n(t)$ are piecewise-constant on each interval $T_m$.  Thus, any result for the arrival model can also be applied to this counting model. Note that, in contrast, it is usually impossible to apply counting results to the general arrival model because driving the interval widths to zero results in vanishing counts within each interval \cite{raginsky_compressed_2010}.

In \cite{jiang_data-dependent_2015}, a variation of the LASSO estimator (similar to that used in~\cite{hansen_lasso_2015}) is proposed that is suitable for solving the Poisson counting inference problem.  Performance bounds for constrained/penalized maximum-likelihood estimators are provided in \cite{raginsky_compressed_2010,rohban_minimax_2016,soni_estimation_2014}.  Of course, these bounds apply only to Poisson counting problems -- nevertheless, since our results also imply bounds for the counting model, it is instructive to compare our bounds with the existing work in this domain. Among the prior work which is most relevant is that of~\cite{rohban_minimax_2016}, from which we specifically draw inspiration in portions of our analysis.  We provide a detailed comparison of our results with those in~\cite{rohban_minimax_2016} in Section~\ref{sec:compare}.

\subsection{Organization of the paper}
We begin in Section~\ref{sec:rec} with the statement of our main recovery guarantee for the estimation of arrival process parameters and several corollaries. In particular, we consider the special case of Poisson counting processes and contrast with the guarantee provided by \cite{rohban_minimax_2016}. Section~\ref{sec:crlb} compares our result to estimates of the Cram\'{e}r-Rao lower bound, as well as to the error we expect in a simple concrete choice of basis $\gamma(t)$.  Section~\ref{sec:app} briefly presents an example application of these results. In Section~\ref{sec:sim} we explore the practical impact of different forms of regularization via simulations.  Finally, we close in Section~\ref{sec:conc} with a discussion of our results.

\section{Recovery guarantees} \label{sec:rec}

\subsection{Maximum likelihood estimation}

The negative log-likelihood of observing a set of events at coordinates $\tau$ under \eqref{eq:arrivalmodel} is given by \cite{daley_introduction_1988} to be
\begin{equation} \label{eq:nll_int}
\mathcal{L}(\tau|x) = \int_{\mathbb{T}} R_x(t) dt - \sum_{m=1}^M \log R_x(\tau_m).
\end{equation}
for any intensity $R_x(t)$ parameterized by $x$ (e.g., the intensity in \eqref{eq:rate}).
Note that an intensity function $R_x(t)$ must be nonnegative across its entire domain.  This can be enforced by defining $\log(z)=-\infty  \;\forall z\le0$.  Accordingly, we constrain $x$ to live in the set $\mathcal{R} = \{x : \inf_{t\in\mathbb{T}}R_x(t)\ge0\}$, which is a convex set under our model~\eqref{eq:rate}.

The maximum likelihood estimate of the true parameters, constrained to lie within $\mathcal{R}$ intersected with an arbitrary set $\mathcal{X}$, is simply
\begin{equation} \label{eq:recprog}
\widehat{x} = \argmin_{x\in\mathcal{X}\cap\mathcal{R}} \mathcal{L}(\tau|x).
\end{equation}
Given the linear model for $R_x(t)$ presented in~\eqref{eq:rate}, we can simplify~\eqref{eq:nll_int} to the equivalent form (omitting the constant $\int_{\mathbb{T}} g(t) dt$ term) of
\begin{equation} \label{eq:nll}
\mathcal{L}(\tau|x) = b^Tx - 1^T \log(g+Ax) ,
\end{equation}
where $b_n = \int_{\mathbb{T}} \gamma_n(t) dt$, $A_{mn} = \gamma_n(\tau_m)$, $g_m = g(\tau_m)$ for $m = 1, \ldots, M$ and $n = 1, \ldots, N$, and the logarithm is taken element-wise over the vector argument.  Note that $A$ and $g$ are both random, as they depend on the random observations $\tau$.
In the Poisson counting model of \eqref{eq:countingmodel}, we instead define $A = [\gamma_1\ldots\gamma_{M_0}]^T$.  In this case~\eqref{eq:nll_int} can be simplified to
\begin{equation} \label{eq:nlldiscrete}
\mathcal{L}(y|x) = 1^TAx - y^T\log(g+Ax) .
\end{equation}
We substitute this likelihood function, depending on the bin counts $y$ instead of arrival coordinates $\tau$, into \eqref{eq:recprog} to form the corresponding maximum likelihood estimation program.

For convex $\mathcal{X}$, the estimation program \eqref{eq:recprog} is convex (for both the likelihood functions in~\eqref{eq:nll} and in~\eqref{eq:nlldiscrete}). This optimization problem can be efficiently solved by algorithms such as those presented in \cite{harmany_this_2012,tran-dinh_composite_2015}.

\subsection{Parameter estimation error bound} \label{sec:mainbound}

In order to state our main result, we will make use of the definition $\Gamma = \text{Gram}(\gamma(t))$, i.e., the $N\times N$ matrix with entries $\Gamma_{ij} = \int_{\mathbb{T}} \gamma_i(t)\gamma_j(t)dt$.  Let $\Tr(\Gamma)$ and $\sigma(\Gamma)$ represent the trace and minimum eigenvalue of $\Gamma$, respectively, and let $\|\gamma\|_{2,\infty}=\sup_{t\in\mathbb{T}} \|\gamma(t)\|_2$.

With these definitions, we can state our main result as follows:

\begin{theorem} \label{thm:main}
If events $\tau$ are produced by a Poisson arrival process with intensity $R_{\bar{x}}(t) \in \{\{0\}\cup [R_{\text{\emph{min}}},R_{\text{\emph{max}}}]\}$ and $R_{\text{\emph{min}}}>2\zeta\frac{\|\gamma\|_{2,\infty}^2}{\sigma(\Gamma)}$, any vector $\widehat{x}$ satisfying $R_{\widehat{x}}(t) \in [0,R_{\text{\emph{max}}}]$ and $\mathcal{L}(\tau|\widehat{x})\le \mathcal{L}(\tau|\bar{x})$ will also satisfy
\begin{equation} \label{eq:thmmain}
\|\widehat{x}-\bar{x}\|_2 < c \frac{\sqrt{\zeta\Tr(\Gamma)}}{\sigma(\Gamma)} \frac{R_{\text{\emph{max}}}}{\sqrt{R_{\text{\emph{min}}}}}
\end{equation}
for an absolute constant $c$ with probability at least $1-(2N+1)\exp(-\zeta)$.
\end{theorem}
The core idea behind the proof (which we borrow from \cite{rohban_minimax_2016}) is to establish the strong convexity of the negative log-likelihood over the domain of our estimator (specifically, $\widehat{x}$ such that $R_{\widehat{x}}(t) \in [0,R_{\text{\emph{max}}}]$).  With strong convexity, we can show that any large deviation from the true solution (i.e., $\widehat{x}$ far from $\bar{x}$) must necessarily violate $\mathcal{L}(\tau|\widehat{x})\le \mathcal{L}(\tau|\bar{x})$, disqualifying it from being the maximizer of the likelihood.  Because the likelihood function depends on a random set $\tau$, this requires the use of concentration inequalities.
The complete proof is detailed in the Appendix.

The parameter $\zeta$ primarily serves as a tradeoff parameter between the scaling of the error and the probability that our bound does not hold, although it also has a role in regulating the minimum $R_\text{min}$ that we can tolerate.  Note that the probability bound is only nontrivial when $\zeta \gtrsim \log N$, thus the theorem contains a dependence on $\log N$ in both the error bound and $R_\text{min}$. In general, however, our primary interest is less on the asymptotic dependence on $N$ and more on the dependence of the error on the structural properties of the underlying dictionary, so we leave this dependence implicit.

\subsection{Estimation under sparsity assumptions}

In many practical settings, especially when the dimension $N$ of the basis $\gamma(t)$ is large, one might expect $\bar{x}$ to be sparse (i.e., to have relatively few nonzeros).  There is a natural extension of Theorem~\ref{thm:main} that can result in an improved bound when this occurs.  To state this result, we will define $\gamma_S(t)$ to be the $|S|$-dimensional basis obtained by taking the basis elements $\gamma_n(t)$ indexed by $S$ and $\Gamma_S$ to be the $|S|\times|S|$ submatrix composed of the rows and columns of $\Gamma$ indexed by $S$ (i.e., the Gram matrix of $\gamma_S(t)$).  Let $\Tr(\Gamma_S)$ and $\sigma(\Gamma_S)$ represent the trace and minimum eigenvalue of $\Gamma_S$, respectively, and let $\|\gamma_S\|_{2,\infty}=\sup_{t\in\mathbb{T}} \|\gamma_S(t)\|_2$.

\begin{theorem} \label{thm:sparserec}
If events $\tau$ are produced by a Poisson arrival process with intensity $R_{\bar{x}}(t) \in \{\{0\}\cup [R_{\text{\emph{min}}},R_{\text{\emph{max}}}]\}$ and $R_{\text{\emph{min}}}>2\zeta\frac{\|\gamma_S\|_{2,\infty}^2}{\sigma(\Gamma_S)}$, any vector $\widehat{x}$ satisfying $\supp(\widehat{x}-\bar{x})\subseteq S$, $R_{\widehat{x}}(t) \in [0,R_{\text{\emph{max}}}]$, and $\mathcal{L}(\tau|\widehat{x})\le \mathcal{L}(\tau|\bar{x})$ will also satisfy
\begin{equation} \label{eq:thmsparse}
\|\widehat{x}-\bar{x}\|_2 < c \frac{\sqrt{\zeta\Tr(\Gamma_S)}}{\sigma(\Gamma_S)} \frac{R_{\text{\emph{max}}}}{\sqrt{R_{\text{\emph{min}}}}}
\end{equation}
for an absolute constant $c$ with probability at least $1-(2|S|+1)\exp(-\zeta)$.
\end{theorem}
The proof is detailed in the Appendix.

Note that the bound in Theorem~\ref{thm:sparserec} depends on the specific choice of index set $S$.  A reader familiar with the sparse approximation literature may wonder how this compares with more traditional results, which hold for arbitrary sparsity patterns, and how the quantities $\Tr(\Gamma_S)$ and $\sigma(\Gamma_S)$ compare to more familiar quantities.  In fact, it is straightforward to restate this result in terms of an appropriate analogue of the {\em restricted isometry property} (RIP)~\cite{davenport_introduction_2012}.
Specifically, let $\mathcal{W} : \mathbb{R}^N \to L_2(\mathbb{T}) $ denote the operator defined by $\mathcal{W}(x) = x^T \gamma(t)$. Observe that we can write $\| \mathcal{W}(x) \|_{L_2(\mathbb{T})}^2 = x^T\Gamma x$.  We say that the operator $\mathcal{W}$ (or equivalently, the basis $\gamma(t)$) satisfies the RIP if there exists a constant $\delta_s \in [0,1)$ such that
\begin{equation*}
(1-\delta_s)\|x\|_2^2 \le x^T\Gamma x \le (1+\delta_s)\|x\|_2^2
\end{equation*}
for all $x$ with at most $s$ nonzeros.

Using this definition, if $\gamma(t)$ satisfies the RIP then we have the bounds $\max_{S : |S|\le s}\Tr(\Gamma_S) \le s(1+\delta_s)$ and $\min_{S : |S|\le s}\sigma(\Gamma_S) \ge 1-\delta_s$.
These bounds let us ignore the specifics of $\supp(\widehat{x}-\bar{x})$ if we can instead bound the \emph{size} of the support.
Accordingly, the bound~\eqref{eq:thmsparse} from Theorem~\ref{thm:sparserec} can be replaced with
\begin{equation} \label{eq:thmrip}
\|\widehat{x}-\bar{x}\|_2 < c \frac{\sqrt{\zeta s(1+\delta_s)}}{1-\delta_s} \frac{R_{\text{max}}}{\sqrt{R_{\text{min}}}} .
\end{equation}

Observe that the results of Theorem~\ref{thm:sparserec} (as stated above or combined with bounds based on the RIP) are really only of interest when we can ensure that the set $\supp(\widehat{x}-\bar{x})$ is sufficiently small.
By the triangle inequality, $|\supp(\widehat{x}-\bar{x})| \le |\supp(\widehat{x})|+|\supp(\bar{x})|$.  Thus, if $|\supp(\widehat{x})|$ and $|\supp(\bar{x})|$ are both small, we can use the RIP of $\gamma(t)$ to obtain an improved error bound.
Unfortunately, enforcing a constraint $\mathcal{X}=\{x : |\supp(x)|<\eta\}$ leads to a challenging nonconvex optimization problem. An open question concerning~\eqref{eq:recprog} is whether we can guarantee that $\widehat{x}$ is sparse when using a sparsity-inducing convex constraint such as $\mathcal{X} = \{ x : \|x\|_1 \le \eta\}$.  We leave such analysis for future work.

\subsection{Guarantees for counting processes}

When using the Poisson counting model \eqref{eq:countingmodel}, the special case of Theorem~\ref{thm:main} follows immediately from the substitutions discussed in Section~\ref{ref:countingsub}.  Specifically, if we define $\|A\|_{2,\infty} = \max_m \sqrt{\sum_n A_{mn}^2}$, $\|A\|_F$ to be the Frobenius norm of $A$, and $\sigma(A)$ to be the smallest singular value of $A$ (the minimum eigenvalue of $\sqrt{A^TA}$), then we have the following result.

\begin{corollary} \label{thm:counting}
Let $y\sim\text{Poisson}(g+A\bar{x})$ with $g+A\bar{x} \in \{\{0\}\cup [R_{\text{\emph{min}}},R_{\text{\emph{max}}}]\}^{M_0}$.  If $R_{\text{\emph{min}}}>2\zeta\frac{\|A\|_{2,\infty}^2}{\sigma^2(A)}$, any vector $\widehat{x}$ satisfying $g+A\widehat{x} \in [0,R_{\text{\emph{max}}}]^{M_0}$ and $\mathcal{L}(y|\widehat{x})\le \mathcal{L}(y|\bar{x})$ will also satisfy
\begin{equation}
\|\widehat{x}-\bar{x}\|_2 < c \frac{\sqrt{\zeta}\|A\|_F}{\sigma^2(A)} \frac{R_{\text{\emph{max}}}}{\sqrt{R_{\text{\emph{min}}}}}
\end{equation}
for an absolute constant $c$ with probability at least $1-(2N+1)\exp(-\zeta)$.
\end{corollary}

We note that this result can easily be extended to the sparse setting in a manner analogous to Theorem~\ref{thm:sparserec} by replacing $\Gamma$ with $\Gamma_S$, $\gamma(t)$ with $\gamma_S(t)$, $A$ with $A_S$, $N$ with $|S|$, and adding the constraint $\supp(\widehat{x}-\bar{x})\subseteq S$.

\subsection{Improving the guarantee via regularization} \label{sec:noisebound}

We note that $g(t)$ has no impact on the result of Theorem~\ref{thm:main}.  This opens the door to what seems like a rather unorthodox strategy for improving this bound.  Specifically, the dependence on the dynamic range $\frac{R_{\text{max}}}{R_{\text{min}}}$ in Theorem~\ref{thm:main} can be removed by simply augmenting our observations with additional events $\rho$ generated by a homogeneous Poisson process with intensity $R_{\text{max}}$.  These additional events will increase the intensity of our process, but in a manner completely known to us.  Specifically, in this case the intensity of $\tau \cup \rho$ is
\begin{equation} \notag
\widetilde{R}_{\bar{x}}(t) = R_{\bar{x}}(t)+R_{\text{max}} = (g(t)+R_{\text{max}}) + \bar{x}^T \gamma(t).
\end{equation}
Note that if $R_{\bar{x}}(t) \in [0,R_{\text{max}}]$ then $\widetilde{R}_{\bar{x}}(t) \in [R_{\text{max}},2R_{\text{max}}]$ and $\frac{\widetilde{R}_{\text{max}}}{\widetilde{R}_{\text{min}}} \le 2$.  Thus, by simply replacing $g(t)$ with $g(t)+ R_{\text{max}}$ in~\eqref{eq:nll} and applying Theorem~\ref{thm:main}, we obtain the following corollary:

\begin{corollary} \label{thm:noised}
If events $\tau$ are produced by a Poisson arrival process with intensity $R_{\bar{x}}(t) \in [0,R_{\text{\emph{max}}}]$, events $\rho$ are produced by a homogeneous Poisson process with intensity $R_{\text{\emph{max}}}$, and $R_{\text{\emph{max}}}>2\zeta\frac{\|\gamma\|_{2,\infty}^2}{\sigma(\Gamma)}$, then any vector $\widehat{x}$ satisfying $R_{\widehat{x}}(t) \in [0,R_{\text{\emph{max}}}]$ and $\mathcal{L}(\tau\cup\rho|\widehat{x})\le \mathcal{L}(\tau\cup\rho|\bar{x})$ will also satisfy
\begin{equation}  \label{eq:thmnoised}
\|\widehat{x}-\bar{x}\|_2 < 2c \frac{\sqrt{\zeta\Tr(\Gamma)}}{\sigma(\Gamma)} \sqrt{R_{\text{\emph{max}}}}
\end{equation}
for an absolute constant $c$ with probability at least $1-(2N+1)\exp(-\zeta)$.
\end{corollary}

To be clear, when we augment our observations $\tau$ with the set of events $\rho$ (with known intensity $\beta$), the negative log-likelihood function that results (ignoring constant terms) is
\begin{equation} \notag
\begin{split}
\iftwocol{\MoveEqLeft}
\mathcal{L}(\tau\cup\rho|x) =  \int_{\mathbb{T}} R_x(t) dt - \sum_m \log (\beta+R_x(\tau_m)) \iftwocol{\\&}
- \sum_\ell \log(\beta + R_x(\rho_\ell)) .
\end{split}
\end{equation}

Comparing the bound in~\eqref{eq:thmnoised} with that in~\eqref{eq:thmmain}, we see that in exchange for an additional factor of two, we have completely removed the dependence on the dynamic range. Moreover, we have replaced the constraint on $R_{\text{min}}$ in Theorem~\ref{thm:main} with a more relaxed constraint on $R_{\text{max}}$.  One might be justifiably skeptical that this will lead to improved performance in practice -- we are essentially adding a {\em large} amount of noise to the signal we wish to estimate.  We explore this empirically in Section~\ref{sec:sim}.  Here, we note that a potentially more palatable strategy might be to perform a deterministic form of regularization that is similar in spirit. Specifically, define the regularized negative log-likelihood
\begin{equation} \label{eq:detreg}
\begin{split}
\iftwocol{\MoveEqLeft}
\mathcal{L}'_\beta(\tau|x) = b^Tx - \sum_m \log\left(\beta+R_x(\tau_m)\right) \iftwocol{\\&}
- \beta\int_{\mathbb{T}} \log\left(\beta+R_x(t)\right)dt ,
\end{split}
\end{equation}
noting that this is equivalent to \eqref{eq:nll} when $\beta=0$.  $\mathcal{L}'_\beta(\tau|x)$ is the result of replacing $\sum_\ell \log(\beta + R_x(\rho_\ell))$ from $\mathcal{L}(\tau\cup\rho|x)$ with its expectation.\footnote{The difficulty of computing the integral in \eqref{eq:detreg} in closed-form means that, typically, numerical integration would be necessary when performing estimation using the regularized likelihood in~\eqref{eq:detreg}.  The use of Riemann integration reduces this result to a sampled version of an arrival model, i.e., a counting model. Recovery as in Corollary~\ref{thm:noised} does not suffer this drawback, remaining comparable to standard maximum likelihood estimation as featured in Theorem~\ref{thm:main}.}

Unfortunately, there does not appear to be an easy extension of our current analysis techniques that would allow us to provide an error bound for the estimate obtained by minimizing~\eqref{eq:detreg}. We conjecture that a result similar to Corollary~\ref{thm:noised} should also hold when optimizing the regularized negative log-likelihood in~\eqref{eq:detreg} with $\beta = R_{\text{max}}$.

The same regularization scheme can be applied to Theorem~\ref{thm:sparserec} and Corollary \ref{thm:counting}, yielding the same consequence of exchanging the dynamic range for a factor of two.

\subsection{Practical considerations}

\subsubsection{Choosing $\mathcal{X}$}
If $\bar{x}\in\mathcal{X}$ then the program in \eqref{eq:recprog} will result in an estimate $\widehat{x}$ that, by construction, satisfies the $\mathcal{L}(\tau|\widehat{x})\le \mathcal{L}(\tau|\bar{x})$ condition of our theorems.
Possible sets $\mathcal{X}$ can include unconstrained vectors, sparse vectors, vectors with limited $\ell_p$ norms, or nonnegative vectors.
Any convex $\mathcal{X}$ will maintain the convexity of \eqref{eq:recprog} -- although convexity is not necessary if we are not concerned with computational efficiency or can still ensure that $\mathcal{L}(\tau|\widehat{x})\le \mathcal{L}(\tau|\bar{x})$ through some alternative argument that does not rely on solving~\eqref{eq:recprog} to global optimality. This results in a tradeoff between constraint sets $\mathcal{X}$ that enable simple efficient algorithms and $\mathcal{X}$ that more tightly enforce desired properties in our recovery $\widehat{x}$. For example, when sparsity is exploited (as in Theorem~\ref{thm:sparserec}) the set $\mathcal{X}=\{x : \|x\|_0\le \|\bar{x}\|_0\}$ (where we define $\|x\|_0=|\supp(x)|$ to count the number of nonzeros in $x$) is a natural choice to ensure a fixed sparsity level in $\widehat{x}$, but results in a nonconvex (and thus challenging) optimization problem. In contrast, $\mathcal{X}=\{x : \|x\|_1\le \eta \}$ is convex and encourages sparse solutions, but does not necessarily guarantee a specific sparsity level.

We also note that our Theorems depend on the condition that $R_{\widehat{x}}(t)\in[0,R_{\text{max}}]$. This can be satisfied by choosing $\mathcal{X}\subseteq\{x:0\le R_x(t)\le R_{\text{max}}\}$.
While the exact value of $R_{\text{max}}$ will likely be unknown, it can be estimated or bounded to sufficient accuracy from $\tau$.  Alternatively, it can be bounded as $R_{\text{max}}\le\|\gamma\|_{2,\infty}\|\bar{x}\|_2$ when we have knowledge of the norm of $\bar{x}$, although this bound is typically poor.  As a matter of practice, we believe that there is often minimal risk in ignoring this constraint completely.  Its role in the proof is solely to limit the pointwise difference $\sup_{t\in\mathbb{T}} |R_{\bar{x}}(t)-R_{\widehat{x}}(t)|\le R_{\text{max}}$, but we expect that a suitable $\widehat{x}$ will accomplish this (at least up to a constant factor) even in the absence of an explicit constraint.

\subsubsection{Sample complexity}
A common lens through which we can evaluate a statistical estimation problem concerns the number of observations required to obtain an accurate estimate of the quantities of interest.  Here our situation is somewhat outside the standard framework as the parameters $\bar{x}$ which we wish to estimate actually determine the number of observations (events) we will obtain.  Nevertheless, we can gain some insight by approaching the problem from this perspective.

For the purpose of generality, we will analyze the number of observations with respect to Theorem~\ref{thm:sparserec}.  In the dense case of Theorem~\ref{thm:main}, simply set $S=\{1,\ldots, N\}$.
It is necessary that $\zeta>\log(2|S|+1)$ for the bound to hold with nonzero probability.  Because $|S|\sigma(\Gamma_S) \le \Tr(\Gamma_S) \le |\mathbb{T}|\|\gamma_S\|_{2,\infty}^2$, we have that
\begin{equation} \notag
\frac{\|\gamma_S\|_{2,\infty}^2}{\sigma(\Gamma_S)}
\ge\frac{|S|}{|\mathbb{T}|}.
\end{equation}
Thus, the requirement that $R_{\text{min}}>2\zeta\frac{\|\gamma_S\|_{2,\infty}^2}{\sigma(\Gamma_S)}$ implies that it is necessary that $|\mathbb{T}|R_{\text{min}} \gtrsim |S|\log|S|$
for the bound to hold with nonzero probability.
Alternatively, in the regularized case of Corollary~\ref{thm:noised} we replace the requirement on $R_{\text{min}}$ with a requirement on $R_{\text{max}}$, resulting in the condition that $|\mathbb{T}|R_{\text{max}} \gtrsim |S|\log|S|$.

On the surface, neither of these conditions directly addresses the issue of how many observations are required.  One may have noticed that the number of events observed, $M=|\tau|$ (or even its expectation~$\overbar{M}$) was conspicuously absent from our theorems.
While not explicitly included, this quantity is implicitly present through the inequalities $|\mathbb{T}|R_{\text{min}}\le\overbar{M}\le|\mathbb{T}|R_{\text{max}}$. Thus when the dynamic range is modest, these conditions essentially reduce to~$\overbar{M}\gtrsim|S|\log|S|$.  This mirrors typical results in the standard setting of sparse recovery with Gaussian statistics \cite{davenport_introduction_2012}.  In non-sparse recovery, we are left with $\overbar{M}\gtrsim N\log N$.
While it is possible that the $\log N$ factor here is only an artifact of our proof (the $N$ factor is shared by Gaussian estimation), it may also be the case that it is necessary to account for the variability of $|\tau|$.

\subsection{Comparison to existing results} \label{sec:compare}

Here we summarize the result of \cite{rohban_minimax_2016} in terms of the quantities introduced here.  The result of \cite{rohban_minimax_2016} holds when $A$, $\widehat{x}$, and $\bar{x}$ are constrained to be nonnegative and $\|\bar{x}\|_0\le s$ holds for some choice $s$.
To make the comparison, we will need to provide several additional definitions from \cite{rohban_minimax_2016} (relabeled to better-reflect our notation):
\begin{align*}
\sigma_*^2(A) &= \min_{\substack{S:|S|\le s,\\x:\|x_S\|_1\ge\|x-x_S\|_1}} \frac{\|Ax\|_2^2}{\|x\|_2^2}
\\R_{\text{min}}^* &= \min_{\substack{x : |x\|_0=s,\\\|x\|_1=\|\bar{x}\|_1}} H(g+Ax)
\\R_{\text{max}}^* &= \|g\|_\infty+\|A\|_{\text{max}}\|\bar{x}\|_1
\end{align*}
where $H(\cdot)$ denotes the harmonic mean of the input vector and $\|\cdot\|_{\text{max}}$ denotes the maximum-magnitude entry of a matrix.  The quantity $\sigma_*^2(A)$ denotes the restricted eigenvalue of $A$ of order $s$ and is roughly comparable to $\min_{|S|\le s} \sigma^2(A_S)$.  Additionally, $R_{\text{min}}^*$ is roughly on the order of $R_{\text{min}}$.  Finally, note that $R_{\text{max}}^*\ge R_{\text{max}}$ and $\sqrt{s M_0}\|A\|_{\text{max}}\ge\|A_S\|_F$ (for any $|S|\le s$) and that these inequalities are often quite loose.

With these definitions, the main result of \cite{rohban_minimax_2016} can be written as
\begin{equation*}
\|\widehat{x}-\bar{x}\|_2 \le \iftwocol{} 54\left(3+\log\tfractwocol{R_{\text{max}}^*}{R_{\text{min}}}\right) \frac{\sqrt{\zeta s M_0}\|A\|_{\text{max}}}{\sigma_*^2(A)} \frac{R_{\text{max}}^*}{\sqrt{R_{\text{min}}^*}}
\end{equation*}
with probability at least $1-2\exp(-\zeta)$ when $\zeta\le\frac{M_0R_{\text{min}}}{H(g+A\bar{x})}\min\{1,R_{\text{min}}\}$.
We remark that this is extremely similar to the findings of the sparse variant of Corollary~\ref{thm:counting}.  However, Corollary~\ref{thm:counting} removes the restriction that $A$, $\widehat{x}$, and $\bar{x}$ be nonnegative. Further, Theorem~\ref{thm:sparserec} generalizes this result to the case of Poisson arrival processes.  We have also presented Corollary~\ref{thm:noised}, which removes the dependence on $R_{\text{min}}$ or $R_{\text{min}}^*$ entirely.  Although it was not the primary aim of this work, we have also considerably improved the dependence on the terms $R_{\text{max}}^*$ and $\|A\|_{\text{max}}$.  However, we also stress that the results in~\cite{rohban_minimax_2016} do have an advantage when dealing with sparse $\bar{x}$ in that their guarantees hold for the (convex) $\ell_1$ constrained estimate (as opposed to requiring the nonconvex sparsity constraint required to obtain a similar guarantee via Theorem~\ref{thm:sparserec}).

\section{Optimality} \label{sec:crlb}

Here we briefly discuss the optimality of our results.  In particular, we will first consider a concrete choice of $\gamma(t)$ for which the expected performance is easy to directly calculate and compare this performance to the bounds established above. Next, we will characterize the Cram\'{e}r-Rao lower bound for this problem and compare our results with this bound.

\subsection{Example: A simple orthobasis} \label{sec:identbasis}

Consider the simple orthobasis of
\begin{equation*}
\gamma_n(t) = \left\{\begin{array}{cc} 1 & n-1\le t<n \\ 0 & \text{otherwise.}\end{array}\right.
\end{equation*}
In this case we have $\Gamma=I$ and $\|\gamma(t)\|_{2,\infty} = 1$.  Since this basis is disjoint (no two elements are supported on intersecting intervals), this is really just a collection of independent Poisson estimation problems with $y_m\sim\text{Poisson}(\bar{x}_m)$.  The maximum likelihood estimate is trivially given by
\begin{equation*}
\widehat{x}_n = \sum_m \gamma_n(\tau_m) ,
\end{equation*}
i.e., the number of events falling in that interval. Hence, $\widehat{x}_n\sim\text{Poisson}(\bar{x}_n)$ and the expected squared error is $\mathbb{E}(\widehat{x}_n-\bar{x}_n)^2 = \bar{x}_n$, from which it follows that
\begin{equation*}
\mathbb{E} \|\widehat{x}-\bar{x}\|_2^2 = \sum_{n=1}^N \bar{x}_n .
\end{equation*}

Applying Theorem~\ref{thm:main} to this problem yields the bound
\begin{equation*}
\|\widehat{x}-\bar{x}\|_2^2 \le c^2 \zeta N R_{\text{max}} \frac{R_{\text{max}}}{R_{\text{min}}}
\end{equation*}
with probability at least $1-(2N+1)\exp(-\zeta)$ when $R_{\text{min}}>2\zeta$.  Our high-probability bound differs from the average-case error of this system (up to a constant) only by the dynamic range.  Alternatively, in Corollary~\ref{thm:noised}, we showed how the dynamic range could be removed from our bound by using a modified recovery program.  Specifically, we obtain a guarantee of the form
\begin{equation*}
\|\widehat{x}-\bar{x}\|_2^2 \le 4c^2 \zeta N R_{\text{max}}
\end{equation*}
with probability at least $1-(2N+1)\exp(-\zeta)$ when $R_{\text{max}}>2\zeta$.  In this case, the bound differs from the expectation by only constants and the ratio between the average and maximum intensity $\bar{x}_n$.

\subsection{Cram\'{e}r-Rao lower bound}

The example described above suggests that for a well-conditioned basis $\gamma(t)$, our analysis appears to be relatively tight. Indeed, by comparing our bound to the Cram\'{e}r-Rao lower bound for this problem, we will obtain additional evidence that this is, in fact, the case.

Towards this end, we note that the Fisher information matrix for \eqref{eq:nll} is
\begin{equation*} \label{eq:fisher}
\mathcal{I}_{\bar{x}} = \left. \mathbb{E}\left(\nabla_x^2\mathcal{L}(x)\right) \right|_{x = \bar{x}} = \mathbb{E}\left(A^T\diag(g+A\bar{x})^{-2}A\right) .
\end{equation*}
The precise value of $\Tr(\mathcal{I}_{\bar{x}}^{-1})$ will depend on the problem (on both $\gamma(t)$ and on $\bar{x}$).  However, one can show that $\mathbb{E}\left(A^T\diag(g+A\bar{x})^{-1}A\right)=\Gamma$ (see Appendix~\ref{prf:term2} for details), which provides the semidefinite ordering
\begin{equation*} \label{eq:invfisher}
R_{\text{max}}^{-1}\Gamma \preceq \mathcal{I}_{\bar{x}} \preceq R_{\text{min}}^{-1}\Gamma .
\end{equation*}
From this we obtain
\begin{equation} \label{eq:trinvbounds}
 \Tr(\Gamma^{-1}) R_{\text{min}}  \le \Tr(\mathcal{I}_{\bar{x}}^{-1})\le \Tr(\Gamma^{-1}) R_{\text{max}}.
\end{equation}

The Cram\'{e}r-Rao lower bound states that any unbiased estimate $\widehat{x}$ of $\bar{x}$ must satisfy
\begin{equation*} \label{eq:lbcrlb}
\mathbb{E} \|\widehat{x}-\bar{x}\|_2^2 \ge \Tr(\mathcal{I}_{\bar{x}}^{-1}).
\end{equation*}

Because the maximum-likelihood estimator asymptotically achieves the Cram\'{e}r-Rao bound under this model \cite{ogata_asymptotic_1978}, \eqref{eq:trinvbounds} implies that, asymptotically, the MLE will satisfy a bound of the form
\begin{equation} \label{eq:asmpytotic}
\mathbb{E} \|\widehat{x}-\bar{x}\|_2^2 \le \Tr(\Gamma^{-1}) R_{\text{max}} .
\end{equation}
There is limited room for improvement here because another asymptotic bound implied by \eqref{eq:trinvbounds} is that
\begin{equation} \notag
\mathbb{E} \|\widehat{x}-\bar{x}\|_2^2 \ge \Tr(\Gamma^{-1}) R_{\text{min}} ,
\end{equation}
as any opposing result would violate the Cram\'{e}r-Rao bound.

Compare these bounds to the result of Corollary~\ref{thm:noised}, which states that (with high probability) the MLE can achieve
\begin{equation} \label{eq:cor1simple}
\|\widehat{x}-\bar{x}\|_2^2 \lesssim \frac{\Tr(\Gamma)}{\sigma(\Gamma)^2} R_{\text{max}}.
\end{equation}
Contrasting~\eqref{eq:asmpytotic} with~\eqref{eq:cor1simple}, we observe that~\eqref{eq:asmpytotic} is a slightly stronger guarantee (ignoring the fact that~\eqref{eq:cor1simple} holds with high probability while~\eqref{eq:asmpytotic} holds only in expectation).  In particular, it is a straightforward consequence of the eigendecomposition of $\Gamma$ that
\begin{equation} \notag
\Tr(\Gamma^{-1}) \le  \frac{\Tr(\Gamma)}{\sigma(\Gamma)^2},
\end{equation}
with equality if and only if the spectrum of $\Gamma$ is flat (i.e., all eigenvalues are equal to each other).  In the case where the basis is ill-conditioned and $\Gamma$ has one or more eigenvalues which are very small compared to the remainder, the bound in~\eqref{eq:asmpytotic} can be somewhat tighter than that in~\eqref{eq:cor1simple}.  However, we note that unless $N$ is quite large, both $\Tr(\Gamma^{-1})$ and $ \frac{\Tr(\Gamma)}{\sigma(\Gamma)^2}$ can both be dominated by $\frac{1}{\sigma(\Gamma)}$, and so when the basis is very poorly conditioned both bounds are rather poor. In total, these results suggest that (up to a constant), there is relatively little room for improvement over the bound in Corollary~\ref{thm:noised}.

\section{Regularization in practice} \label{sec:sim}

Here we take a practical look into the claims of Corollary~\ref{thm:noised} and our conjecture concerning our deterministically regularized alternative.  Specifically, Corollary~\ref{thm:noised} provides improved theory (at least when $R_{\text{min}}$ is small) when we augment our observations with additional random events (i.e., noise). We would like to understand if such gains are actually to be expected in practice.

\begin{figure}[t]
\centering
\includegraphics[width=0.42\textwidth]{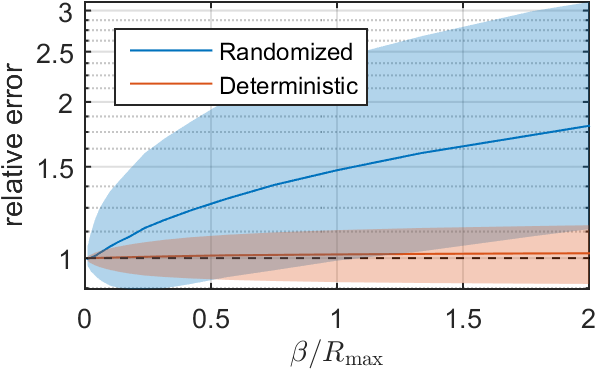}
\caption{Error $\|\widehat{x}-\bar{x}\|_2$ with regularization of intensity $\beta$ compared to error without regularization.  Lines represent the median relative error and the shaded region denotes the space between the 10\% and 90\% quantiles.  Values less than one denote improvement over unregularized recovery while values above one denote degraded performance.}
\label{fig:reg}
\end{figure}

Towards this end, we choose a basis of $N=50$ shifted-Gaussian basis functions, sampled as to produce a counting process with $M_0=500$ observations ($A\in\mathbb{R}^{M_0\times N}$).  We choose coefficient vectors $\bar{x}$ with entries that are independent and identically exponentially distributed such that the expected number of events is $\overbar{M}=100$.  We then simulate the Poisson process with the intensity given by this basis and these parameters and compute the $\ell_2$-norm error between the true $\bar{x}$ and maximum likelihood estimate $\widehat{x}$.  We compare this error to the error of computing the regularized estimator using random (Corollary~\ref{thm:noised}) or deterministic (using \eqref{eq:detreg}) regularization.  Figure~\ref{fig:reg} plots this result for different choices of regularization parameter $\beta$ relative to $R_{\text{max}}$ over many trials.  Note that Corollary~\ref{thm:noised} uses $\beta=R_{\text{max}}$, and larger $\beta$ denotes stronger regularization (and $\beta=0$ denotes no regularization).  Relative error larger than 1 denotes degraded performance with regularization and less than 1 denotes improved performance.

While we only include simulations for a single choice of basis and parameters, these results are representative of the behavior we observed for every choice of basis/parameters we have attempted. Examining Figure~\ref{fig:reg}, we note that randomized regularization (Corollary~\ref{thm:noised}) frequently results in significantly-degraded performance.  Larger regularization results in larger error as the noise overwhelms the signal.  Deterministic regularization does not suffer the same increasing error with increasing regularization, but instead saturates at error that is, in the median, slightly higher than the unregularized error.  We infer that the effect of regularization is simply to perturb the solution in a way that will sometimes improve the error and will sometimes degrade it, without biasing our solution in some way that results in consistent improvement.

In summary, despite our best efforts, we have not uncovered any regime in which the regularization suggested by Corollary~\ref{thm:noised} (or our conjecture) systematically improved recovery over the unregularized case.
This leaves open the possibility that the bound of Corollary~\ref{thm:noised} may be more representative of actual performance than that of Theorem~\ref{thm:main} (differing by a dependence on the dynamic range of the underlying intensity) even when the prescribed regularization is completely omitted.  In other words, it is possible that the results of Corollary~\ref{thm:noised} actually hold without any regularization at all.
However, our analysis does not provide an obvious means to establish this result.

\section{Conclusions} \label{sec:conc}

In this paper we have provided novel recovery guarantees for parameter estimation in the context of Poisson arrival processes. These guarantees are near-optimal. However, we note that this optimal theoretical performance is established for a version of the MLE which has been augmented with noise -- a procedure that does not appear to actually result in improved performance in practice.

There are several remaining improvements that may yet be possible.  First, the results presented here require that (the unknown part of) the Poisson intensity function lie within the span of the basis.  It may also be possible to address recovery when the true intensity is only well-approximated by this basis. Second,
the existence of Corollary~\ref{thm:noised}, which adds noise to the observations to improve theoretical performance, combined with the findings of Sections~\ref{sec:crlb} and \ref{sec:sim} suggests that the dynamic range dependence in Theorem~\ref{thm:main} may be an artifact of the proof and ultimately unnecessary.  We continue to conjecture that an improved bound should be possible for our deterministically-regularized MLE (or a similar alternative), but leave further exploration of this issue for future work. Finally, many important open questions remain concerning the use of more sophisticated convex regularizers (e.g., an $\ell_1$-norm constraint when $\bar{x}$ is sparse).

\ifnottwocol{
\section*{Acknowledgments}
This work was supported by grants NRL N00173-14-2-C001, AFOSR FA9550-14-1-0342, and NSF CCF-1350616 as well as support from the Alfred P. Sloan Foundation.
}

\bibliographystyle{IEEEtran}
\bibliography{bibabbrevs,pprefs}

\ifnottwocol{
\appendix
\section{Proof of Theorems~\ref{thm:main} and \ref{thm:sparserec}} %\label{sec:proof}
}
\iftwocol{
\appendix[Proof of Theorems~\ref{thm:main} and \ref{thm:sparserec}] %\label{sec:proof}
}

The overall outline of our proof is inspired by the proof of the main result of \cite{rohban_minimax_2016}. We will provide a proof of Theorem~\ref{thm:sparserec} -- Theorem~\ref{thm:main} follows as a special case of the same argument.

Define $\Delta' = \widehat{x}-\bar{x}$.  Our goal is to show that
\begin{equation} \label{eq:proofstart}
\|\Delta'\|_2<\epsilon \qquad \forall \Delta' \text{ s.t. } \mathcal{L}(\tau|\bar{x}+\Delta') < \mathcal{L}(\tau|\bar{x}) .
\end{equation}
A sufficient condition for \eqref{eq:proofstart} is that
\begin{equation} \label{eq:proofgoal}
\mathcal{L}(\tau|\bar{x}+\Delta') - \mathcal{L}(\tau|\bar{x}) \ge 0 \qquad \forall \Delta' \text{ s.t. } \|\Delta'\|_2\ge\epsilon.
\end{equation}
In other words, any large deviation from $\bar{x}$ implies an inferior likelihood to that of $\bar{x}$, and hence any $\widehat{x}$ with a smaller negative log-likelihood than $\bar{x}$ must lie close to $\bar{x}$.

Suppose that $\Delta'$ has $s$ nonzero elements.  We can define a $\Delta\in\mathbb{R}^s$ and a selector matrix $S\in\{0,1\}^{N\times s}$ (a submatrix of the $N \times N$ identity matrix) such that $\Delta' = S\Delta$.  We make the definitions $b_S = S^Tb$, $A_S = AS$, $\Gamma_S = S^T\Gamma S$, and $\gamma_S(t)=S\gamma(t)$.  As a shorthand, we will define $D = \diag(g+A\bar{x})^{-1}$.  The negative log-likelihood gap between the perturbed and true solutions reduces to
\begin{equation*}
\mathcal{L}(\tau|\bar{x}+S\Delta) - \mathcal{L}(\tau|\bar{x}) = b_S^T\Delta - 1^T \log\left(1+DA_S\Delta\right).
\end{equation*}
We use the bound $\log(1+x) \le x - \frac{3x^2}{6+4x}$, along with some simple algebraic manipulation, to obtain
\begin{equation} \label{eq:boundafterlog}
\begin{split}
\iftwocol{\MoveEqLeft}
\mathcal{L}(\tau|\bar{x}+S\Delta) - \mathcal{L}(\tau|\bar{x}) \ge b_S^T\Delta - 1^T DA_S\Delta \iftwocol{\\&}
+ 3\Delta^TA_S^TD\diag(6(g+A\bar{x}) + 4A_S\Delta)^{-1}A_S\Delta .
\end{split}
\end{equation}
Constraining $R_{\bar{x}}(t)$ and $R_{\widehat{x}}(t)$ to the interval $[0,R_{\text{max}}]$ ensures that $\|g+A\bar{x}\|_\infty\le R_{\text{max}}$ and $\|A_S\Delta\|_\infty\le R_{\text{max}}$.  Employing these bounds allows us state that
\begin{equation*}
\diag(6(g+A\bar{x}) + 4A_S\Delta)^{-1} \succeq \frac{1}{10R_{\text{max}}}I
\end{equation*}
and thus we can lower-bound the right-hand side of \eqref{eq:boundafterlog} by
\begin{equation*}
(b_S - A_S^TD1)^T\Delta + \frac{3}{10R_{\text{max}}} \Delta^TA_S^TDA_S\Delta .% \ge 0 \qquad \forall \|\Delta\|_2\ge\epsilon .
\end{equation*}
After applying the Cauchy-Schwarz inequality and basic properties of singular values, we obtain
\begin{equation*}
\begin{split}
\iftwocol{\MoveEqLeft}
\mathcal{L}(\tau|\bar{x}+S\Delta) - \mathcal{L}(\tau|\bar{x}) \ge \iftwocol{\\&}
- \left\|b_S - A_S^T D1\right\|_2 \|\Delta\|_2 + \frac{3\sigma(A_S^T DA_S)}{10R_{\text{max}}} \|\Delta\|_2^2
\end{split}
\end{equation*}
where $\sigma(\cdot)$ denotes the minimum %right
singular value of matrix.  Accordingly,
\begin{equation*}
\|\Delta\|_2 \ge \frac{10R_{\text{max}}\left\|b_S - A_S^T D1\right\|_2}{3\sigma(A_S^T DA_S)}
\end{equation*}
implies that $\mathcal{L}(\tau|\bar{x}+S\Delta) - \mathcal{L}(\tau|\bar{x})\ge0$.  Thus, the choice $\epsilon = {10R_{\text{max}}\left\|b_S - A_S^T D1\right\|_2}/{3\sigma(A_S^T DA_S)}$ is sufficient to satisfy \eqref{eq:proofgoal}.  This expression involves two random quantities, $\|b_S - A_S^T D1\|_2$ and $\sigma(A_S^T DA_S)$.  These are random because $A$ and $D$ depend on the (random) observations $\tau$ of our process.  We address these quantities using the following lemmas:

\begin{lemma} \label{lem:term1}
Using the preceding definitions, the inequality
\begin{equation*}
\left\|b_S - A_S^T D1\right\|_2 \le \frac{2}{3}\frac{\zeta\|\gamma_S\|_{2,\infty}}{R_{\text{\emph{min}}}} + \sqrt{2\zeta\frac{M}{\overbar{M}}\frac{\Tr(\Gamma_S)}{R_{\text{\emph{min}}}}}
\end{equation*}
holds with probability at least $1-(s+1)\exp(-\zeta)$.
\end{lemma}
 Proof: See Appendix~\ref{prf:term1}.

\begin{lemma} \label{lem:term2}
Using the preceding definitions, the inequality
\begin{equation*}
\sigma(A_S^T DA_S) \ge \sigma(\Gamma_S) \left(1-\sqrt{\frac{2\zeta\|\gamma_S\|_{2,\infty}^2}{R_{\text{\emph{min}}}\sigma(\Gamma_S)}}\right)
\end{equation*}
holds with probability at least $1-s\exp(-\zeta)$.
\end{lemma}
Proof: See Appendix~\ref{prf:term2}.

By applying Lemmas~\ref{lem:term1} and \ref{lem:term2} to our expression for $\epsilon$ and applying a union bound, with probability at least $1-(2s+1)\exp(-\zeta)$ the error is bounded
\begin{equation*}
\epsilon \le \frac{10R_{\text{max}}}{3} 
\frac{\frac{2}{3}\frac{\zeta\|\gamma_S\|_{2,\infty}}{R_{\text{min}}} + \sqrt{2\zeta\frac{\Tr(\Gamma_S)}{R_{\text{min}}}}}{\sigma(\Gamma_S) \left(1-\sqrt{\frac{2\zeta\|\gamma_S\|_{2,\infty}^2}{R_{\text{min}}\sigma(\Gamma_S)}}\right)} .
\end{equation*}

If $R_\text{min} \ge \alpha\zeta\frac{\|\gamma_S\|_{2,\infty}^2}{\sigma(\Gamma_S)}$ for some $\alpha$ then, using the inequalities $s\sigma(\Gamma_S)\le\Tr(\Gamma_S)\le|\mathbb{T}|\|\gamma_S\|_{2,\infty}^2$, we bound
\begin{equation*}
\zeta
\le \frac{R_\text{min}\sigma(\Gamma_S)}{\alpha\|\gamma_S\|_{2,\infty}^2}
\le \frac{R_\text{min}\Tr(\Gamma_S)}{\alpha s \|\gamma_S\|_{2,\infty}^2} .
\end{equation*}
By defining $c_{\alpha,s} = \frac{10}{3} \frac{\frac{2}{3\sqrt{\alpha s}}+\sqrt{2}}{1-\sqrt{\frac{2}{\alpha}}}$, it follows that
\begin{equation*}
\epsilon \le c_{\alpha,s} \frac{\sqrt{\zeta\Tr(\Gamma_S)}}{\sigma(\Gamma_S)} \frac{R_\text{max}}{\sqrt{R_\text{min}}}
\end{equation*}
with probability at least $1-(2s+1)\exp(-\zeta)$.

The constant $c_{\alpha,s}$ shrinks as $\alpha$ or $s$ grow (though other terms in the bound grow with $s$).
For any $\alpha>2$, $c_{\alpha,s}$ is well-defined and bounded by an absolute constant for any $s$.  For example, with the trivial choice $s\ge1$ we can bound $c_{3,s}<33$, $c_{5,s}<16$, and $c_{\infty,s}\approx4.71$.
This completes the proof of Theorem~\ref{thm:sparserec}.  Theorem~\ref{thm:main} is a result of the choice $s=N$ (i.e., $S=I$).

\subsection{Proof of Lemma~\ref{lem:term1}} \label{prf:term1} % matrix Bernstein

The elements of the vector $b-A^TD1$ are given by the expression
\begin{equation*}
\left[b - A^T D1\right]_n = b_n - \sum_{m=1}^M \frac{\gamma_n(\tau_m)}{R_{\bar{x}}(\tau_m)} .
\end{equation*}
Recall that the event coordinates $\tau_m$ are independent and identically distributed with probability density function $f_{\tau_m}(t) = \frac{R_{\bar{x}}(t)}{\overbar{M}}$ and that $M\sim\text{Poisson}(\overbar{M})$.  Define a positive integer $Q\ge\overbar{M}$, and $w_q^{(Q)}\sim\text{Bernoulli}(\overbar{M}/Q)$.
Because a Poisson random variable can be formed by taking an infinite sum of Bernoulli random variables (or a Binomial random variable with an infinite number of terms) with finite expectation, $[A^T D1]_n$ has the same distribution as
\begin{equation} \label{eq:bernmult}
\lim_{Q\rightarrow\infty} \sum_{q=1}^Q w_q^{(Q)} \frac{\gamma_n(\tau_q)}{R_{\bar{x}}(\tau_q)}
\end{equation}
when we consider a set $\{\tau\}$ with $Q$ events.

It is easily shown that
\begin{equation*}
\mathbb{E}\left(w_q^{(Q)} \frac{\gamma_n(\tau)}{R_{\bar{x}}(\tau)}\right) = \frac{\overbar{M}}{Q} \int_{\mathbb{T}} \frac{\gamma_n(t)}{R_{\bar{x}}(t)} f_{\tau_m}(t) dt = \frac{b_n}{Q}
\end{equation*}
\begin{equation*}
\mathbb{E}\left(\left(w_q^{(Q)}\frac{\gamma_n(\tau)}{R_{\bar{x}}(\tau)}\right)^2\right) = \frac{\overbar{M}}{Q} \int_{\mathbb{T}} \frac{\gamma_n^2(t)}{R_{\bar{x}}^2(t)} f_{\tau_m}(t) dt \le \frac{\Gamma_{nn}}{QR_{\text{min}}} .
\end{equation*}
Let $S$, when used as an index set, be the indices of the rows of the matrix $S$ with a nonzero entry.  We use the above statements to control the expectations
\begin{equation*}
\mathbb{E}\left(\frac{b}{Q} - w_q^{(Q)} \frac{\gamma(\tau)}{R_{\bar{x}}(\tau)}\right) = 0 ,
\end{equation*}
\begin{equation*}
\sum_{n\in S} \mathbb{E}\left(\left(\frac{b_n}{Q} - w_q^{(Q)}\frac{\gamma_n(\tau)}{R_{\bar{x}}(\tau)}\right)^2\right) %= \sum_{n\in S} \mathbb{E}\left(\left(\frac{\gamma_n(\tau)}{R_{\bar{x}}(\tau)}\right)^2\right) - \frac{b_n^2}{\overbar{M}^2}
\le \frac{\Tr(\Gamma_S)}{QR_{\text{min}}} .%- \frac{\|b_S\|_2^2}{\overbar{M}^2}
\end{equation*}
Further, we bound $\left\|\frac{b_S}{Q} - w_q^{(Q)}\frac{\gamma_S(t)}{R_{\bar{x}}(t)}\right\|_2 \le \left(\tfrac{\overbar{M}}{Q}+1\right)\frac{\|\gamma_S\|_{2,\infty}}{R_{\text{min}}}$.
The matrix Bernstein inequality \cite{tropp_introduction_2015} (applied to a $s\times1$ vector) assures us that
\begin{equation*}
\begin{split}
\iftwocol{\MoveEqLeft}
\mathbb{P}\left(\left\|b_S - \sum_{q=1}^Q w_q^{(Q)}\frac{\gamma_S(\tau_q)}{R_{\bar{x}}(\tau_q)}\right\|_2 \ge \eta \right) \iftwocol{\\&}
\le (s+1)\exp\left(\frac{-\eta^2}{2\frac{\Tr(\Gamma_S)}{R_{\text{min}}} + \frac{2}{3}\eta\left(1+\tfrac{\overbar{M}}{Q}\right)\frac{\|\gamma_S\|_{2,\infty}}{R_{\text{min}}}}\right) .
\end{split}
\end{equation*}
We solve for the argument of the exponent and use the concavity of the square root ($\sqrt{a^2+b^2} \le |a|+|b|$) to write
\begin{equation*}
\begin{split}
\iftwocol{\MoveEqLeft}
\mathbb{P}\left(\left\|b_S - \sum_{q=1}^Q w_q^{(Q)}\tfrac{\gamma_S(\tau_q)}{R_{\bar{x}}(\tau_q)}\right\|_2 \iftwocol{\right.\\&\left.} \ge
\frac{2}{3}\left(1+\frac{\overbar{M}}{Q}\right)\frac{\zeta\|\gamma_S\|_{2,\infty}}{R_{\text{min}}}
+ \sqrt{2\zeta\frac{\Tr(\Gamma_S)}{R_{\text{min}}}}\right)
\iftwocol{\\&} \le (s+1)\exp(-\zeta) .
\end{split}
\end{equation*}
Because $A_S^T D1$ has the same distribution as \eqref{eq:bernmult}, we let $Q\rightarrow\infty$ to complete the proof.

\subsection{Proof of Lemma~\ref{lem:term2}} \label{prf:term2}
As a shorthand, define $G = A^T DA$ and $G^{(m)}_{ij} = \frac{\gamma_i(\tau_m)\gamma_j(\tau_m)}{R_{\bar{x}}(\tau_m)}$ so that $G = \sum_{m=1}^M G^{(m)}$.  With this definition,
\begin{equation*}
\mathbb{E}\left(G^{(m)}_{ij}\right) = \mathbb{E}\left(\frac{\gamma_i(\tau)\gamma_j(\tau)}{R_{\bar{x}}(\tau)}\right) = \frac{\Gamma_{ij}}{\overbar{M}} .
\end{equation*}
$G^{(m)}$ is rank-$1$ and positive semidefinite.  We can write and bound the spectral norm as
\begin{equation*}
\left\|S^TG^{(m)}S\right\|_2 = \frac{\sum_{n\in S} \gamma_n^2(\tau_m)}{R_{\bar{x}}(\tau_m)} \le \frac{\|\gamma_S\|_{2,\infty}^2}{R_{\text{min}}} .
\end{equation*}
As in Appendix~\ref{prf:term1}, we argue that $G$ has the same distribution as $G' = \lim_{Q\rightarrow\infty} \sum_{q=1}^Q w_q^{(Q)} G^{(q)}$ when $w_q^{(Q)}\sim\text{Bernoulli}(\overbar{M}/Q)$.

Applying the matrix Chernoff inequality \cite{tropp_introduction_2015} provides the bound, for $\eta\in[0,1]$,
\begin{equation*}
\begin{split}
\iftwocol{\MoveEqLeft}
\mathbb{P}\left(\sigma(S^TG'S) \le (1-\eta)\sigma(\Gamma_S)\right) \iftwocol{\\&}
\le s\exp\left(-\frac{R_{\text{min}}\sigma(\Gamma_S)\eta^2}{2\|\gamma_S\|_{2,\infty}^2}\right) .
\end{split}
\end{equation*}
The proof is completed by solving for the argument of the exponent and letting $Q\rightarrow\infty$ so that the distributions of $G$ and $G'$ converge.

\end{document}